\documentclass[aps,prd,nofootinbib,superscriptaddress]{revtex4}
\usepackage[sumlimits]{amsmath} 
\usepackage{epsfig}
\usepackage{amssymb}
\usepackage{graphicx}
\usepackage{pstricks}
\usepackage{color}
\usepackage{bbold}
\usepackage{hyperref}

\begin{document}

\title[Article Title]{Conformal metric perturbations and boundary term as physical source}

\author{Juan Ignacio Musmarra} 
\email{jmusmarra@mdp.edu.ar}
\affiliation{Departamento de F\'{i}sica, Facultad de Ciencias Exactas y Naturales, Universidad Nacional de Mar del Plata (UNMdP), Dean Funes 3350, C.P. 7600, Mar del Plata, Argentina \\
Instituto de Investigaciones F\'{i}sicas de Mar del Plata (IFIMAR), Consejo Nacional de Investigaciones Cient\'{i}ficas y T\'{e}cnicas (CONICET), Mar del Plata, Argentina}

\author{Claudia Moreno} 
\email{claudia.moreno@cucei.udg.mx}
\author{Rafael Hern\'{a}ndez-Jim\'{e}nez}
\email{rafaelhernandezjmz@gmail.com}
\affiliation{Departamento de F\'isica,
Centro Universitario de Ciencias Exactas y Ingenier\'ias, Universidad de Guadalajara
Av. Revoluci\'on 1500, Colonia Ol\'impica C.P. 44430, Guadalajara, Jalisco, M\'exico}

%
%
\bigskip
\begin{abstract}
In the context of the Relativistic Quantum Geometry formalism, where the cosmological constant is promoted to a dynamical variable by attributing it a geometric interpretation as a result of a flux on the boundary of a manifold and establishing a connection between the perturbations of the Ricci tensor and the metric tensor, we propose an approach for the perturbations of the metric tensor. From this, imposing that Einstein's equations must hold for the tensors defined from the perturbed quantities obtained from conformal transformations, we derive a functional form for the cosmological parameter $\Lambda$ in terms of the cosmological parameter $\bar{\Lambda}$ of the perturbed manifold. We then use the obtained equations to propose a cosmological model based on the Friedmann-Lema\^{i}tre-Robertson-Walker metric with no spatial curvature, fitting the free parameters using observational data from Hubble and Type Ia Supernovae. The model is statistically comparable to $\Lambda$CDM; although, the joint analysis produces a smaller $H_{0}^{\rm Conformal}=69.80\rm\,\, Km \,s^{-1}\,Mpc^{-1}$ in contrast to the flat $\Lambda$CDM result $H_{0}^{\Lambda\rm CDM}=70.52\rm\,\, Km \,s^{-1}\,Mpc^{-1}$. An evident singularity occurs when the conformal factor $\Xi^{2}=2$, yields an early universe dominated only by matter $\rho_{m}$, which undoubtedly does not correspond to a viable history of our cosmos. Despite these limitations, a specific scenario remains feasible. This study aims to offer insights into the acceleration of the universe and addresses key questions in contemporary cosmology.
\end{abstract}


\maketitle

\section{Introduction}\label{sec1}

With the purpose of pave the way towards a consistent quantum theory of gra\-vi\-ta\-tio\-nal interaction, a plethora of possible modifications to General Relativity have been proposed in the literature~\cite{Arkani-Hamed:2003pdi,Carroll:2004de,Sotiriou:2008rp,Sotiriou:2009xt,DeFelice:2010aj,Olmo:2011uz,Clifton1,deRham:2014zqa,Cai:2015emx,Nojiri:2017ncd,CANTATA:2021asi,Shankaranarayanan1}. These modifications, without entering into tension with observational data, seek to extend the predictive scope of this theory and solve its inconsistencies with Quantum Field Theory. One of these proposals consists of replacing the cosmological constant $\Lambda$ with a cosmological parameter $\Lambda(x^{\alpha},...)$ that can depend on the coordinates $x^{\alpha}$, physical parameters and/or additional fields, thus modifying the conservation equation of the energy-momentum tensor~\cite{Velten:2021xxw,Bonder:2022kdw}. From a theo\-re\-ti\-cal point of view, this idea has been developed in the context of unimodular gravity theories\cite{Chiou:2010ne,Jain:2011jc,Carballo-Rubio:2022ofy,Landau:2022mhm,Alvarez:2023utn,Fabris:2023vop,Jirousek:2023gzr,Kluson:2023yzv,LinaresCedeno:2023cpn}.

There are various physical and geometric arguments to justify replacing the cosmological constant with a dynamical variable~\cite{Saez-Gomez:2011mll,Alexander:2018djy,Faraggi:2020blm,Bernardo:2022cck,Buchmuller:2022msj,Comelli:2023otq}. One of these motivations consists of attributing the existence of a cosmological parameter to a geometric flow on the boun\-da\-ry of a Riemannian manifold, in the context of the Lagrangian formulation of General Relativity~\cite{mb1}. Starting from the Einstein-Hilbert action in a manifold with a boundary, instead of considering the usual treatment of incorporating the Gibbons-Hawking-York counterterm to give rise to a well-defined variational principle~\cite{Poisson:2009pwt}, a link between the perturbations of the metric tensor and the Ricci tensor is proposed, establishing a proportionality between these quantities given by the parameter $\Lambda$~\cite{mb1,Hernandez-Jimenez:2022daw,Bellini:2023ddy,Hernandez-Jimenez:2023vnb,Bellini:2024khn,Musmarra:2019sug,Musmarra:2022qhj,Bellini:2024meo,Bellini:2025snh}:
\begin{equation}\label{ec1}
\delta{R}_{\alpha\beta}=-\Lambda\,\delta{g}_{\alpha\beta}.
\end{equation}
When considering this link in the variational principle, the Einstein equations are obtained with the cosmological constant replaced by an arbitrary dynamic variable, whose functional form would depend on the geometric flow on the boundary of the manifold associated with a certain physical system of interest for study in par\-ti\-cu\-lar. This proposal has led to interesting results from applications in inflationary models~\cite{Bellini:2023ddy,Hernandez-Jimenez:2022daw,Hernandez-Jimenez:2023vnb,Bellini:2025snh} and relativistic thermodynamics~\cite{Musmarra:2019sug,Musmarra:2022qhj,Bellini:2024meo} in the context of the Relativistic Quantum Geometry formalism~\cite{mb1}, where metric perturbations are attributed a quantum origin by proposing that the perturbations $\delta\Gamma^{\alpha}_{\;\;\beta\gamma}$ correspond to the expectation value of an operator $\delta\hat{\Gamma}^{\alpha}_{\;\;\beta\gamma}(\sigma_{\alpha})$ defined from the derivatives of a scalar field $\sigma$. A recent alternative based on this idea consists of assuming that the boundary of the manifold corresponds to a non-Riemannian manifold and assuming a form for the derivative of the metric tensor perturbations based on a Weyl-integrable geometry~\cite{Romero:2012hs,Banados:2024rfy}, which introduces extra terms in the Einstein equations and allows the study of dissipative quintessence~\cite{Matei:2024xsa}.

An observation that can be made about this proposal is that an explicit form for the parameter $\Lambda$ is not established, so in general some ad hoc functional forms have been employed to study the phenomenological effects in certain particular cases. Recently, a functional form for $\Lambda$ has been introduced within the framework of this formalism concerning a parameter $z$ based on geometric motivations, considering modifications to the normalization condition of the four-velocities of relativistic observers~\cite{Bellini:2024khn,Bellini:2025snh}.

In this article, we will motivate a functional form for $\Lambda$ by following a different way. To this end, we make use of conformal transformations, which have been considered in gravitational theories as a way to provide an alternative mathematical description with respect to a classical gravitational theory~\cite{Dabrowski:2008kx}, with the equivalence between these different representations being a subject of debate~\cite{Shapiro:1995kt,Faraoni:1998qx}. Conformal transformations have also been used to construct black hole solutions embedded in cosmological backgrounds, which are relevant for addressing issues related to the loss of predictability in cosmological spacetimes~\cite{Hammad:2018hhv}. Another motivation for studying conformal transformations in General Relativity is that they could be associated with a hidden symmetry from which the cosmological constant might emerge in the dynamical equations~\cite{BenAchour:2020xif,Banados:2024rfy}.

The structure of this article is as follows. In Section 2, we calculate the deviations of a conformal manifold with respect to the background Riemannian manifold to express a particular case of eq.~\eqref{ec1}, and we also impose that Einstein's equations with cosmological parameter $\bar{\Lambda}$ hold on the background manifold. For these two aspects to be consistent with each other, we show that a transformation rule relating $\Lambda$ to $\bar{\Lambda}$ must be satisfied. In Section 3, we present a cosmological framework to illustrate the proposal of conformal perturbations as a physical source of the dynamical equations, where we illustrate the behavior of the cosmological parameter based on the free parameters of this scenario. In Section 4, we perform a statistical fit of the model's free parameters using late-time Observational Hubble data and Type Ia Supernovae distance modulus. In Section 5, some comments are made on the statistical analysis and the results obtained. In Section 6, we present some conclusions of the work carried out and mention aspects to be considered that could be addressed in future research.

\section{Cosmological parameter from conformal perturbations}\label{sec2}

In this section, we will begin by introducing the formalism presented in~\cite{mb1} and subsequent works, and then propose a particular case in a specific context. We will start by considering the Einstein-Hilbert action ${\cal I}_{EH}$ in the metric formalism:
\begin{equation}\label{EH}
{\cal I}_{EH}=\int_{\mathcal{M}} d^4x \,\sqrt{-g} \left[ \frac{R}{2\kappa} + {\cal L}\right],
\end{equation}
where $\kappa=8\pi G/c^4$, and $\mathcal{M}$ is a manifold equipped with the metric tensor $g_{\alpha\beta}$ and the Levi-Civita connection. By varying with respect to the perturbations of the metric tensor and considering the principle of least action, we obtain:
\begin{equation}\label{variacionEH}
\delta {\cal I}_{EH} = \int d^4 x \sqrt{-g} \left[ \delta g^{\alpha\beta} \left( G_{\alpha\beta} - \kappa T_{\alpha\beta}\right)
+ g^{\alpha\beta} \delta R_{\alpha\beta} \right]=0,
\end{equation}
where the stress-energy tensor is defined as
\begin{equation}
T_{\alpha\beta}=\frac{-2}{\sqrt{-g}}\frac{\delta(\sqrt{-g}\mathcal{L})}{\delta{g}^{\alpha\beta}}.
\end{equation}
Instead of incorporating counterterms into the action to yield a well-defined variational principle, we incorporate the link eq.~\eqref{ec1} into eq.~\eqref{variacionEH} to obtain the Einstein equations with cosmological parameter:
\begin{equation}\label{Einsteinecmods}
G_{\alpha\beta}+\Lambda\,g_{\alpha\beta}=\kappa\,T_{\alpha\beta}.
\end{equation}
Moreover, this border term can be absorbed into the stress energy tensor, providing an additional energy component in the full description. Thus, the covariant derivative 
\begin{equation}\label{covariant_D}
\nabla_{\beta}\,T^{\alpha\beta} = \frac{1}{\kappa}\,g^{\alpha\beta}\nabla_{\beta}\Lambda \,,
\end{equation}
yields a physical-sourced equation of energy conservation. This means that the flux due to the boundary term becomes the source of the matter sector.

The approach of~\cite{mb1} consists of constructing a dynamic equation that takes into account the contribution of boundary terms by combining eqs.~\eqref{variacionEH} and \eqref{Einsteinecmods}:
\begin{equation}
\Lambda\,g^{\alpha\beta}\,\delta{g}_{\alpha\beta}+g^{\alpha\beta}\,\delta{R}_{\alpha\beta}=0,
\end{equation} 
while also considering the following assumptions:
\begin{enumerate}
\item The perturbations occur over an extended manifold $\bar{\mathcal{M}}$, characterized by deviations with respect to the Levi-Civita connection $\delta\bar{\Gamma}^{\alpha}_{\;\;\beta\gamma}$.
\item The perturbations $\delta{g}_{\alpha\beta}$ occur in the direction of the velocity of a relativistic observer $U^{\gamma}$: $\delta{g}_{\alpha\beta}=g_{\alpha\beta|\gamma}\,U^{\gamma}$, where $|$ denotes the covariant derivative with respect to the connection on the extended manifold $\bar{\Gamma}^{\alpha}_{\;\;\beta\gamma}=\Gamma^{\alpha}_{\;\;\beta\gamma}+\delta\bar{\Gamma}^{\alpha}_{\;\;\beta\gamma}$.
\item The perturbations $\delta{R}_{\alpha\beta}=\delta\Gamma^{\lambda}_{\;\;\alpha\beta;\lambda}-\delta\Gamma^{\lambda}_{\;\;\alpha\lambda;\beta}$ can be replaced in eq.~\eqref{variacionEH} by perturbations over the extended manifold $\delta{R}_{\alpha\beta}=\delta\bar{\Gamma}^{\lambda}_{\;\;\alpha\beta|\lambda}-\delta\bar{\Gamma}^{\lambda}_{\;\;\alpha\lambda|\beta}$.
\end{enumerate}

In this article, we will use a different approach to develop eq.~\eqref{ec1}. Instead of assuming that the perturbations of the metric tensor are proportional to $U^{\alpha}$, we will impose that the $\delta{g}_{\alpha\beta}$ are linked to conformal deviations with respect to a solution of Einstein's equations:
\begin{equation}\label{perturbacionesg}
    \delta{g}_{\alpha\beta}=\bar{g}_{\alpha\beta}-g_{\alpha\beta}=(\Xi^2-1)\,g_{\alpha\beta}, \quad \bar{g}_{\alpha\beta}=\Xi^2\,g_{\alpha\beta}.
\end{equation}
From this, the rest of the metric perturbations defined with respect to $\delta{g}_{\alpha\beta}$ can be computed. In particular,  
\begin{equation}\label{deltaRicci}
    \delta{R}_{\alpha\beta}=\bar{R}_{\alpha\beta}-R_{\alpha\beta}=\frac{1}{\Xi^2}\left(4\,\Xi_{;\alpha} \Xi_{;\beta}-\Xi_{;\gamma} \,\Xi^{;\gamma} \,g_{\alpha\beta}\right)-\frac{1}{\Xi}\left(2\,\Xi_{;\alpha\beta}+g_{\alpha\beta}\,\Box\Xi\right),
\end{equation}
and by introducing eq.~\eqref{deltaRicci} into eq.~\eqref{ec1} and taking the trace with respect to $g^{\alpha\beta}$, an expression relating the conformal factor $\Xi$ to the cosmological parameter $\Lambda$ can be obtained:
\begin{equation}\label{onda1}
    \Box\Xi=\frac{2}{3}\,\Lambda\,\Xi\,\left(\Xi^2-1\right).
\end{equation}

An important contribution of this new approach is that it allows one to determine a functional form for the cosmological parameter $\Lambda$, which in~\cite{mb1} and subsequent works is generally treated as an arbitrary quantity, dependent on a particular physical system and associated with a physically meaningful flux on the boundary of the manifold $\mathcal{M}$: $\delta\Phi=g^{\alpha\beta}\,\delta R_{\alpha\beta}=\Lambda\,g_{\alpha\beta}\,\delta{g}^{\alpha\beta}$. With this motivation, we propose that from $\bar{g}_{\alpha\beta}$, a manifold $\bar{\mathcal{M}}$ can be defined where Einstein's equations hold:
\begin{equation}\label{Einsteinconfs}
    \bar{G}_{\alpha\beta}+\bar{\Lambda}\,\bar{g}_{\alpha\beta}=\kappa\,\bar{T}_{\alpha\beta}, \quad \bar{G}_{\alpha\beta}=\bar{R}_{\alpha\beta}-\frac{1}{2}\,\bar{R}\,\bar{g}_{\alpha\beta}, \quad \bar{T}_{\alpha\beta}=\frac{{T}_{\alpha\beta}}{\Xi^2},
\end{equation}
where $\bar{\Lambda}$ corresponds to a transformation of $\Lambda$ that ensures the invariance of Einstein's equations under the transformations $\{g_{\alpha\beta} \rightarrow \bar{g}_{\alpha\beta},\,\Lambda \rightarrow \bar{\Lambda}\}$. From this, the perturbations of the Einstein tensor are given by
\begin{equation}
    \delta{G}_{\alpha\beta}=\bar{G}_{\alpha\beta}-G_{\alpha\beta}=\Lambda\,g_{\alpha\beta}-\bar{\Lambda}\,\bar{g}_{\alpha\beta}+k\,\delta{T}_{\alpha\beta},
\end{equation}
where $\delta{T}_{\alpha\beta}=\bar{T}_{\alpha\beta}-T_{\alpha\beta}$, and through algebraic manipulations and considering \eqref{ec1}, we obtain
\begin{equation}\label{onda2}
    \Box\Xi=\frac{\Xi^3}{3}\left(\Lambda-\bar{\Lambda}\right)+\frac{k\,(\Xi^2-1)\,T}{12\,\Xi^2},
\end{equation}
where $T = T_{\alpha\beta} g^{\alpha\beta}$, so for this result to be consistent with eq.~\eqref{onda1}, a direct comparison between eqs.~\eqref{onda1} and \eqref{onda2} leads to the conclusion that the transformation for the cosmological parameter must be given by
\begin{equation}\label{translambda}
    \Lambda\rightarrow\bar{\Lambda}=\left(\frac{2-\Xi^2}{\Xi^2}\right)\Lambda+\Lambda_{T}, \quad \Lambda_{T}=\frac{k\,(\Xi^2-1)\,T}{4\,\Xi^5},
\end{equation}
where, as expected, $\Lambda = \bar{\Lambda}$ when $\Xi^2 = 1$. In this way, the cosmological parameter $\Lambda$ can be expressed in terms of the conformal factor $\Xi$, the trace of the energy-momentum tensor $T$, and the parameter $\bar{\Lambda}$. These quantities, in turn, define an arbitrary conformal manifold where Einstein's equations with a cosmological parameter hold. In particular, among all possible transformations, we will consider a manifold where $\bar{\Lambda} = \Lambda_0 + \Lambda_T$, where $\Lambda_0$ corresponds to the cosmological constant of the standard cosmological model. From this, the cosmological parameter $\Lambda$, which determines the dynamical eq.~\eqref{Einsteinecmods}, is established through the equations:
\begin{equation}\label{onda3}
    \Lambda=\frac{\Xi^2}{2-\Xi^2}\Lambda_0, \quad \Box\Xi=\frac{2\,\Xi^3\,(\Xi^2-1)}{3\,(2-\Xi^2)}\Lambda_0.
\end{equation}

\section{A cosmological model with $\Lambda(z)$}\label{cosmological_model}
To illustrate this proposal of conformal perturbations as a physical source of the dynamical equations, we will consider a manifold described by a flat homogeneous and isotropic Friedmann-Lema\^{i}tre-Robertson-Walker (FLRW) metric:
\begin{equation}\label{frw}
ds^2=-dt^2+a(t)^2\left[dr^2+r^2(d\theta^2+\sin^2(\theta)d\phi)\right] \,,
\end{equation}
where $t$ is the cosmological time, $a=a(t)$ is the scale factor. With the energy-momentum tensor corresponding to an ideal fluid. From this, the dynamic equations are given by:
\begin{eqnarray}
&& \hspace{-2cm} 3H^2=\rho_{m} + \rho_{rad} +\Lambda \,, \label{H2}\\
&& \hspace{-2cm} 2\dot{H}+3H^2= - (p_{rad} + p_{m})  + \Lambda \,, \label{dotH} \\
\dot{\rho}_{rad} + 3H\rho_{rad} + 3p_{rad}H = 0 \,, && 
\dot{\rho}_{m} + 3H\rho_{m} + 3p_{m} H = -\dot{\Lambda} \,, \label{dotrho}
\end{eqnarray}
where $H=\dot{a}/a$ is the expansion rate or the Hubble parameter. We have assumed that the cosmological parameter interacts solely with the matter sector. The matter density $\rho_{m}$ is the sum of baryonic matter $\rho_{b}$ and cold dark matter $\rho_{cdm}$. Radiation density comprises photons $\rho_{\gamma}$ and ultra-relativistic neutrinos $\rho_{\nu}$, so $\rho_{rad}=\rho_{\gamma}+\rho_{\nu}$. In a barotropic component, the state equation is $p=\omega \rho$, with $\omega$ being the barotropic parameter. Baryons ($b$) and cold dark matter ($cdm$) act as dust, having zero pressure, thus $p_{m}=0$. Radiation follows $p_{rad} = \rho_{rad}/3$. Additionally, we will assume that $\Xi=\Xi(z)$, here \( z \) is the redshift, so that:
\begin{equation}\label{lambdaz}
\Lambda=\Lambda(z)=\frac{\Xi(z)^2}{2-\Xi(z)^2}\Lambda_0 \,,
\end{equation}
at $z=0$ we take $\Xi(z=0)=1$ so $\Lambda(z=0)=\Lambda_{0}$ (the cosmological constant), thus restoring the usual $\Lambda$CDM model. Furthermore, the normalized Hubble parameter can be written as follows:
\begin{equation}\label{HubbleNorm1}
E^{2}(z) \equiv \frac{H^{2}(z)}{H_{0}^{2}} =  \Omega_{0\,rad}\left(1+z\right)^{4} + \Omega_{0\,m}f(z,\Theta) +  \Omega_{0\,\Lambda}\left(\frac{\Xi(z)^2}{2 - \Xi(z)^2}\right)  \,,
\end{equation}
where $\Omega_{0\,i}=\rho_{0\,i}/(3H_{0}^{2})$ ($i=rad,\, m$), $\Omega_{0\,\Lambda}=\Lambda_{0}/(3H_{0}^{2})$; also $f(z,\Theta)=\rho_{m}/\rho_{0\,m}$ is the solution of $\rho_{m}=\rho_{m}(z,\Theta)$, and $\Theta$ represents the free parameter space. We will apply the normalized Friedmann equation at $z=0$, which means $E^{2}(z=0)=1$, to derive specific parameters such as $\Omega_{0\,\Lambda}$. Then, in a flat FLRW universe, the Friedman constraint is satisfied during all cosmological evolution, that is, $\sum_{i}\Omega_{i}(z)=1$, where
\begin{eqnarray}
&& \Omega_{rad}(z)=\frac{\Omega_{0\,rad}\left(1+z\right)^{4}}{E^{2}(z)} \,,\,  
\Omega_{m}(z) =\frac{\Omega_{0\,m}f(z,\Theta)}{E^{2}(z)} \nonumber\\ && \hspace{2cm} \Omega_{\Lambda}(z) = \frac{\Omega_{0\,\Lambda}}{E^{2} (z)}\left(\frac{\Xi(z)^2}{2 - \Xi(z)^2}\right),
\end{eqnarray}
and to measure the cosmic acceleration of the expansion of the universe, we use the deceleration parameter, which is given by the relation $q=-\ddot{a}a/\dot{a}^{2}=-1-\dot{H}/H^{2}$, where in terms of the redshift, and $E(z)$ is given by:
\begin{equation}
q = \frac{1}{E^{2}(z)}\left[\Omega_{0\,rad}\left(1+z\right)^{4} + \frac{\Omega_{0\,m}}{2}f(z,\Theta) - \Omega_{0\,\Lambda}\left(\frac{\Xi(z)^2}{2-\Xi(z)^2}\right) \right] \,.     
\end{equation}
Then, by taking the time derivative $\Lambda(z)$ (eq.~\ref{lambdaz}) and formulas $(a/a_{0})=(1+z)^{-1}$ hence $dt=-(1+z)^{-1}H^{-1}dz$, the conservation of energy equation becomes:
\begin{equation}
\frac{df(z,\Theta)}{dz} = \frac{3f(z,\Theta)}{1+z} - \frac{4\Omega_{0}\Xi}{\left(2 - \Xi^2\right)^{2}}\frac{d\Xi}{dz}   \,, 
\end{equation}
here $\Omega_{0}=\Omega_{0\,\Lambda}/\Omega_{0\,cdm}$; and using the variable $f(z,\Theta)=\rho_{m}/\rho_{0\,m}$, so initially $f(z=0,\Theta)=1$. On the other hand, the evolution equation of the conformal factor $\Xi$ (eq.~(\ref{onda3})) is given by:
\begin{eqnarray}
&& (1+z)^{2}E^{2}(z)\frac{d^{2}\Xi}{dz^{2}}+(1+z)\left[\frac{3}{2}\Omega_{0\,m}f(z,\Theta) + 2\Omega_{0\,rad}\left(1+z\right)^{4}  - 2E^{2}(z)\right]\frac{d\Xi}{dz} \nonumber \\
&& \hspace{5cm} = \frac{2\Xi^{3}(\Xi^{2}-1)}{(\Xi^{2}-2)}\Omega_{0\,\Lambda} \,.
\end{eqnarray}
To address the preceding equation, we employ a numerical algorithm. First, we simplify the system at the present cosmological time, and since the contribution $\Omega_{0\,rad}\ll 1$ is not considered initially. Specifically, for $z=0$, we take $E(z=0) = 1$, $\Xi(z=0)=1$, and $f(z=0,\Theta) = 1$, resulting in a simplified equation:
\begin{equation}
(1+z)\frac{d^{2}\Xi}{dz^{2}} +  \left[\frac{3}{2}\Omega_{0\,m} - 2\right]\frac{d\Xi}{dz} = 0 \,,  
\end{equation}
and its solution is: 
\begin{equation}
\Xi = c_{1} + \frac{8(z+1)^{3}(2z+2)^{-3\Omega_{0\,m}/2}}{3(2-\Omega_{0\,m})}c_{2} \,.    
\end{equation}
Note that in order to be consistent we must take $c_{1}=1 - (2^{3-3\Omega_{0\,m}/2})/(3(2-\Omega_{0\,m}))c_{2}$ so at $z=0$ we have $\Xi(z=0)=1$. On the other hand, its derivative $d\Xi/dz(z=0)=2^{2-3\Omega_{0\,m}/2}c_{2}$ will be used as an initial condition to numerically solve the whole system at any $z$. So, this constant $c_{2}$ will be fixed by observational data presented in the next section. 

\begin{figure}[htbp] 
\includegraphics[scale=0.75]{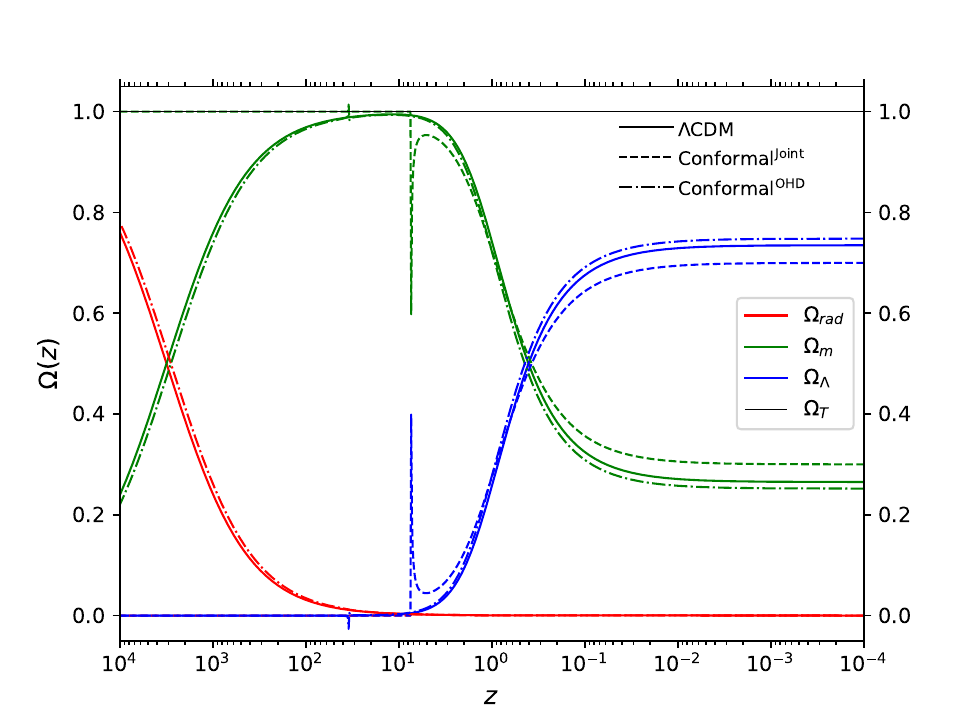}
\caption{Cosmological evolution of $\Omega_{\Lambda}(z)$ (blue), $\Omega_{m}(z)$ (green), $\Omega_{rad}(z)$ (red), and $\Omega_{T}(z)=\sum_{i}\Omega_{i} = 1$ (black). Solid lines correspond to ($\Lambda$CDM) case; dashed ones for Conformal$^{\rm Joint}$ (Supernovae Ia + Observational Hubble Data) model with $c_{2}=0.0079$; and dashed dots Conformal$^{\rm OHD}$ (Observational Hubble Data) with $c_{2}=0.0008$. Both vertical cuts represent the divergence for $\Xi^{2}=2$. In the Conformal$^{\rm Joint}$ case the anomaly occurs at $z\sim 7$; while in the Conformal$^{\rm OHD}$ instance its irregularity is at $z\sim 34$.}\label{fig:Omegas}
\end{figure}

\begin{figure}[htbp] 
\includegraphics[scale=0.75]{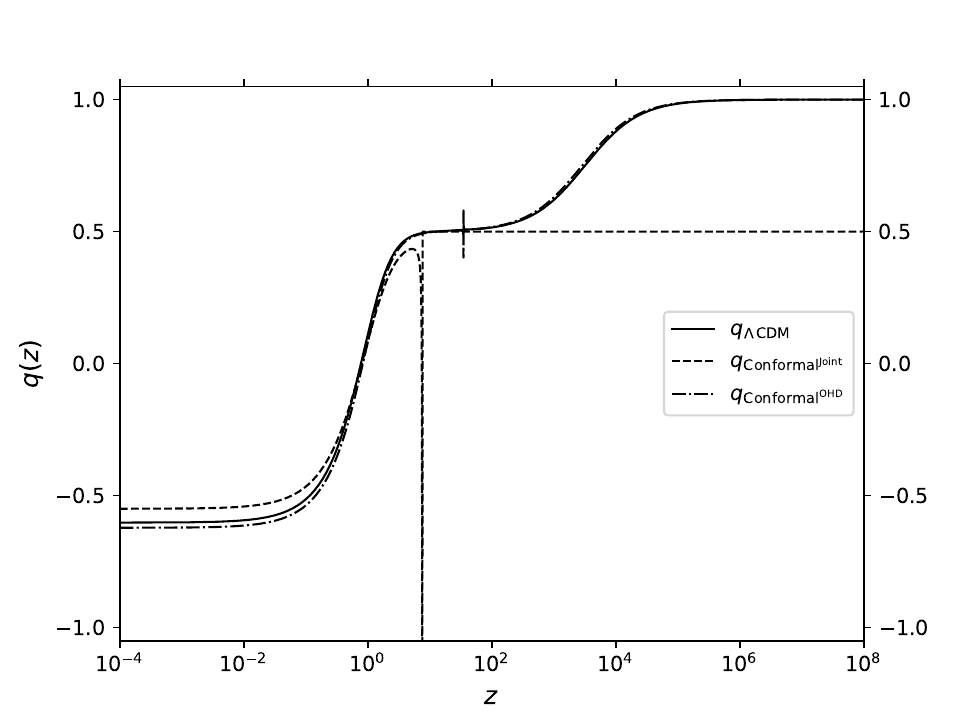}
\caption{Evolution of the deceleration parameter $q(z)$ in terms of the redshift $z$. Solid lines correspond to $\Lambda$CDM framework; then dashed ones for Conformal$^{\rm Joint}$ (Supernovae Ia + Observational Hubble Data) model with $c_{2}=0.0079$; and dashed dots Conformal$^{\rm OHD}$ (Observational Hubble Data) with $c_{2}=0.0008$. Both vertical cuts represent the divergence for $\Xi^{2}=2$. In the Conformal$^{\rm Joint}$ case the anomaly occurs at $z\sim 7$; while in the Conformal$^{\rm OHD}$ instance its irregularity is at $z\sim 34$.}\label{fig:q}
\end{figure}

Now, we present an example taking the following input numerical values due to our statistical analysis performed in the next section (Sect.~\ref{analysis_results}): for $\Lambda$CDM: $\Omega_{0\,rad}=8.39\times 10^{-5}$, $\Omega_{0\,m}=0.265$, and $\Omega_{0\,\Lambda}=1-\Omega_{0\,rad}-\Omega_{0\,m}=0.735$; for Conformal$^{\rm OHD}$ (Observational Hubble Data): $\Omega_{0\,rad}=8.19\times 10^{-5}$, $\Omega_{0\,m}=0.252$, $\Omega_{0\,\Lambda}=1-\Omega_{0\,rad}-\Omega_{0\,m}=0.748$, and $c_{2}=0.0008$; for Conformal$^{\rm Joint}$ (Supernovae Ia + Observational Hubble Data): $\Omega_{0\,rad}=8.57\times 10^{-5}$, $\Omega_{0\,m}=0.300$, $\Omega_{0\,\Lambda}=1-\Omega_{0\,rad}-\Omega_{0\,m}=0.700$, and $c_{2}=0.0079$. Figs.~\ref{fig:Omegas} and~\ref{fig:q} show the behaviour of all $\Omega(z)$'s and $q(z)$, respectively. We have chosen the groups of outcomes that come from two observational data sets. 


Let us first address the elephant in the room. The dynamical equation of the cosmological parameter $\Lambda$ (eq.~\ref{lambdaz}) contains in its denominator a contingent divergence for $\Xi^{2}=2$, this oddity presents an issue that can be avoided, at least numerically, by implementing a numeric artifact, say including a $\epsilon \sim 10^{-20}$ in the denominator. This ensures that all differential equations can be computed without any problematic substantial error; in fact, whilst $\Xi^{2}$ does not reach the value 2, there is no miscalculation. However, as soon as $\Xi^{2} = 2$ a mistake occurs in the numerical implementation, leading to unavoidable errors. Although this approach circumvents any divergence, irremediable falls into some distinct issues observed on some relevant parameters, such as all $\Omega(z)$'s and $q(z)$. For instance, in both figs.~\ref{fig:Omegas} and~\ref{fig:q} there are two abrupt vertical lines, where the divergences occur, i.e. at $\Xi^{2} = 2$. However, these two conformal scenarios show quite different evolutions for $z\gg 1$. In the Conformal$^{\rm Joint}$ case the anomaly occurs at $z\sim 7$; while in the Conformal$^{\rm OHD}$ instance its irregularity is at $z\sim 34$. Notwithstanding, both scenarios reveal singularities, the Conformal$^{\rm Joint}$ example displays a universe dominated completely by the matter component ($\rho_{m}$), right after its divergence, leaving radiation absolutely erased from its past, which undoubtedly does not correspond to a viable history of our cosmos. This result is even more evident in the deceleration parameter, yielding $q(z)=2$, which corresponds to a world dominated only by matter. On the other hand, the conformal$^{\rm OHD}$ result yields a completely different story. All $\Omega(z)$'s and $q(z)$ do follow expected evolution, they present similar behaviours to the standard $\Lambda$CDM framework. In conclusion, despite the fact that both results exhibit irregularities, the larger $z$ the divergence occurs, which does not severely affect all relevant parameters. 


%
%

%
%
\section{Cosmological constraints}\label{cosmo-constraints}
In this section, we constrain the free parameters of the model $\Theta =\{h, \Omega_{0\, m}, c_{2} \}$, where $h=H_{0}/100$. To achieve this task, a merit function $\log \mathcal{L} \sim$ $\chi^2$ is minimised by using late-time Observational Hubble Data (OHD) and Type Ia Supernovae (SNe Ia) distance modulus. Then, we compute the best fit value by means of the affine-invariant Markov Chain Monte Carlo method (MCMC) \cite{2010CAMCS...5...65G}. To compute posterior probabilities, we use Cobaya software \cite{Torrado:2020dgo}, which is a general-purpose Bayesian analysis code. Note that we estimate the free parameters and their confidence regions via a Bayesian statistical analysis; then we apply the Gaussian likelihood function:
\begin{equation}\label{distributions}
\mathcal{L}_I \sim \exp\left(-\frac{\chi^2_I}{2} \right)\,,
\end{equation}
here, $I$ stands for each data set under consideration, namely OHD, SNe Ia; and their joint analysis with $\chi^{2}_{\rm joint}=\chi^{2}_{\rm OHD}+\chi^{2}_{\rm SNe}$. They are described in the following segments, along with their corresponding $\chi^{2}$ functions. We compute $\Omega_{0\,rad}=2.469\times 10^{-5}h^{-2}(1+0.2271\,N_{eff})$ \cite{2011ApJS..192...18K, Magana:2017nfs}, where $N_{eff}=3.04$ is the standard number of relativistic species \cite{Mangano:2001iu}. Also, from the normalised Friedmann equation at $z=0$, that is, $E^{2}(z=0)=1=\Omega_{0\,rad} + \Omega_{0\,m} + \Omega_{0\,\Lambda}$, we obtain $\Omega_{0\,\Lambda}=1-\Omega_{0\,rad}-\Omega_{0\,m}$. Moreover, since we have no previous knowledge about the parameters to analyse, we have considered flat priors, since they are the most conventional to use. They are $\Omega_{0\, m}\in [0.1,0.5]$; $h\in [0.5,0.8]$; and $c_{2}\in [-0.03,0.03]$. Furthermore, to monitor the convergence of the posteriors, we employ the Gelman–Rubin criterion~\cite{10.2307/2246093}, $R-1$. We have selected $R-1<1.0\times 10^{-2}$ for both the $\Lambda$CDM and CBFP scenarios. 
%
%
\subsection{Observational Hubble Data}
We calculate the optimal model parameter, $H_0$, by minimising the merit function:
\begin{equation}\label{chiOHD}
\chi_{\rm OHD}^2=\sum_{i=1}^{N_H} \left(\frac{H_{th}(z_i, \Theta(h, \Omega_{0\, m}, \lambda  ))-H_{obs}(z_i)}{\sigma_{obs}(z_{i})} \right)^2 \,,
\end{equation}
where $H_{th}$ and $H_{obs}$ are the theoretical and observational Hubble parameters at redshift $z_{i}$, respectively; then $\sigma_{obs}(z_{i})$ is the associated error of $H_{obs}(z_{i})$; and $\Theta(h,\Omega_{0\, m},c_{2})$ denotes the free parameter space of $H_{th}$ (eq.~\eqref{HubbleNorm1}). The sample consists of $N_{H}=52 \, H(z)$ measurements in the redshift range $0.0 < z< 2.36$ \cite{Magana:2017nfs}. These data comes from Baryon Acoustic Oscillations (BAO)~\cite{2011MNRAS.416.3017B, 10.1093/mnras/stx721, 10.1093/mnras/sty506, deSainteAgathe:2019voe, Blomqvist:2019rah} and Cosmic Chronometers~\cite{Jimenez:2001gg}. 
%
%
\subsection{Supernovae Ia: SNe Ia}
Data from SNe Ia observations are usually released as a distance modulus $\mu$. In our study, we will use the compilation of observational data for $\mu$ given by the Pantheon Type Ia catalog \cite{Pan-STARRS1:2017jku}, which consists of $N_{\mu}=1048$ SNe data samples, which includes observations up to redshift $z=2.26$. The model for the observed distance modulus $\mu$ is~\cite{Pan-STARRS1:2017jku, Kessler:2016uwi}: 
\begin{equation}\label{mu_proposed}
\mu = m_{b} - \mathcal{M} \,,
\end{equation}
where $m_b$ is the apparent B-band magnitude of a fiducial SNe Ia, and $\mathcal{M}$ is a nuisance parameter, which in fact is strongly degenerated with respect to $H_{0}$ \cite{Pan-STARRS1:2017jku}. To overcome this problem, we will follow the BEAMS method proposed in \cite{Pan-STARRS1:2017jku, Kessler:2016uwi}. First, the theoretical distance modulus in a flat FLRW geometry is given by: 
\begin{equation}
\mu_{th}(z_i, \Theta)=5\log_{10}\left(\frac{d_L(z_i, \Theta)}{Mpc}\right)+\bar{\mu} \,, \quad \bar{\mu}=5\left[\log_{10}{\left(c\right)}+5\right]
\end{equation}
where $c$ is the speed of light given in units of $\rm km\,s^{-1}$, and $d_L(z_i, \Theta)$ is the luminosity distance: 
\begin{equation}
d_L(z_i, \Theta)=\frac{(1+z_{i})}{H_0}\int_0^{z_{i}}\frac{dz'}{E(z',\Theta)} \,.
\end{equation}
This relation allows us to contrast our theoretical model with respect to the observations by minimising the merit function:
\begin{equation}\label{chiSNe}
\chi_{\rm SNe}^2=\sum_{i=1}^{N_{\mu}} \left(\frac{\mu_{th}(z_i, \Theta)-\mu_{i}}{\sigma_{i}} \right)^2 \,,
\end{equation}
where $\mu_i$ and $\mu_{\rm th}$ are the observational and theoretical distance modulus of each SNe Ia at redshift $z_i$, respectively; $\sigma_i$ is the error in the measurement of $\mu_i$, and $\Theta=\{h, \Omega_{0\, m}, c_{2} \}$ represents all the free parameters of the respective model. However, we will follow another method to simplify our analysis. First, eq.~\eqref{chiSNe} can be written in matrix notation (bold symbols): 
\begin{equation}\label{chiSNematrix}
\chi^{2}_{\rm SNe} = \mathbf{M}^\dagger\mathbf{C}^{-1}\mathbf{M},
\end{equation}
where $\mathbf{C}$ is the total covariance matrix given by:
\begin{equation}\label{covariancematrix}
\textbf{C}=\textbf{D}_{\rm stat}+\textbf{C}_{\rm sys} \,, 
\end{equation}
and $\mathbf{M}=\mathbf{m}_{b}-\mbox{\boldmath$\mu$}_{\rm th}\left(z_{i},\theta\right)-\mbox{\boldmath$\mathcal{M}$}$ \cite{Corral:2020lxt}. The diagonal matrix $\textbf{D}_{\rm stat}$ only contains the statistical uncertainties of $m_{b}$ for each redshift, whilst $\textbf{C}_{\rm sys}$ denotes the systematic uncertainties in the BEAMS with the bias correction approach. We can even simplify this method by reducing the number of free parameters and marginalising over $\mathcal{M}$, we use $\mathcal{M}=\bar{\mathcal{M}}-\bar{\mu}$ with $\bar{\mathcal{M}}$ being an auxiliary nuisance parameter~\cite{Corral:2020lxt}. Moreover, eq.~(\ref{chiSNematrix}) can be expanded as~\cite{Lazkoz:2005sp,Corral:2020lxt}:
\begin{equation}\label{chi2projected}
\chi^{2}_{\rm SNe}=A\left(z,\theta\right)-2B\left(z,\theta\right)\bar{\mathcal{M}}+C\bar{\mathcal{M}}^{2} \,, 
\end{equation}
where
\begin{equation}
A\left(z,\theta\right) = \bar{\mathbf{M}}^{\dagger}\textbf{C}^{-1}\bar{\mathbf{M}} \,,\quad B\left(z,\theta\right) = \bar{\mathbf{M}}^{\dagger}\textbf{C}^{-1}\,\textbf{1} \,,\quad C = \textbf{1}^{\dagger}\, \textbf{C}^{-1}\, \textbf{1} \,, 
\end{equation}
with $\bar{\mathbf{M}}=\mathbf{m}_{B}-\mbox{\boldmath$\mu$}_{\rm th}\left(z_{i},\theta\right)+\bar{\mbox{\boldmath$\mu$}}$~\cite{Corral:2020lxt}. Note that, in fact, $\bar{\mathbf{M}}$ no longer contains any troublesome parameters. Finally, minimising eq.~\eqref{chi2projected} with respect to $\bar{\mathcal{M}}$ gives $\bar{\mathcal{M}}=B/C$ and reduces to:
\begin{equation}\label{chi2SNeprojected}
\chi^{2}_{\rm SNe}\Big|_{\rm min}=A\left(z,\theta\right)-\frac{B\left(z,\theta\right)^{2}}{C}. 
\end{equation}
Both eqs.~(\ref{chi2SNeprojected}, \ref{chiSNematrix}) yield the same information; however, eq.~(\eqref{chi2SNeprojected}) only contains the free parameters of the model, and the nuisance term has been marginalised. Thus, we will implement eq.~(\ref{chi2SNeprojected}) as our merit function. The Pantheon data set is available online in the GitHub repository \href{https://github.com/dscolnic/Pantheon}{https://github.com/dscolnic/Pantheon}: the document \textit{lcparam\_full\_long.txt} contains the corrected apparent magnitude $m_{b}$ for each SNe Ia together with their respective redshifts ($z_{i}$) and errors ($\sigma_{i}$); and the file \textit{sys\_full\_long.txt} includes the full systematic uncertainties matrix $\textbf{C}_{\rm sys}$.
%
%
\section{Analysis and Results}\label{analysis_results}
Data from SNe Ia alone yields bias results due to the nuisance parameter $\mathcal{M}$, therefore, $H_{0}$ cannot be determined using only this set of information. Therefore, to constrain $H_{0}$, we must combine it with other observations. Both $\Lambda$CDM and Conformal models are contrasted with a Joint analysis using OHD and SNe Ia data through their corresponding Hubble parameters. Once the model is constrained, we will compare the proposed Conformal model with the $\Lambda$CDM one using the Bayesian Information Criterion (BIC) \cite{BIC} defined as:
\begin{equation}
BIC = \chi_{min}^2+k \ln{N} \,,
\end{equation}
where $\chi_{min}^2$ is the log-likelihood of the model, $k$ is the number of free parameters of the optimised model; and $N$ is the number of data samples. This criterion gives us a quantitative value to select among several models. Following Jeffrey-Raftery's \cite{10.2307/271063} guidelines, if the difference in BICs between the two models is 0–2, this constitutes `weak' evidence in favour of the model with the smaller BIC; a difference in BICs between 2 and 6 constitutes `positive' evidence; a difference in BICs between 6 and 10 constitutes `strong' evidence; and a difference in BICs greater than 10 constitutes `very strong' evidence in favour of the model with smaller BIC.

\begin{figure}[htbp] 
\includegraphics[scale=0.85]{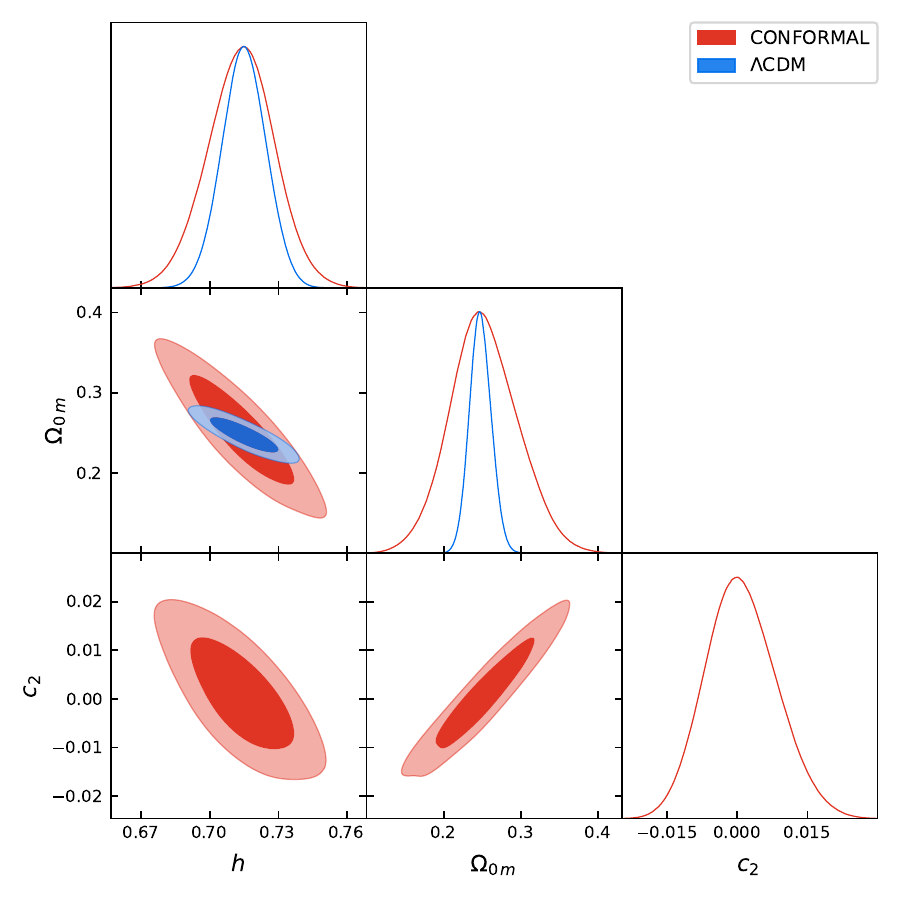}
\caption{OHD data constraints of $h$, $\Omega_{0\,m}$, and $c_{2}$ for the flat $\Lambda$CDM model and the Conformal scenario, using Bayesian statistical analysis of Sec.~\ref{cosmo-constraints}. The admissible regions correspond to $1\sigma\left(68.3\%\right)$, and $2\sigma\left(95.5\%\right)$, CL, respectively. In this case, the best-fit values of the Conformal scenario at $1\sigma$ are $h=0.714^{+0.014}_{-0.014}$, $\Omega_{0\,m}=0.252^{+0.041}_{-0.046}$, and $c_{2}=0.0008^{+0.0071}_{-0.0080}$.}\label{OHD}
\end{figure}

\begin{figure}[htbp] 
\includegraphics[scale=0.85]{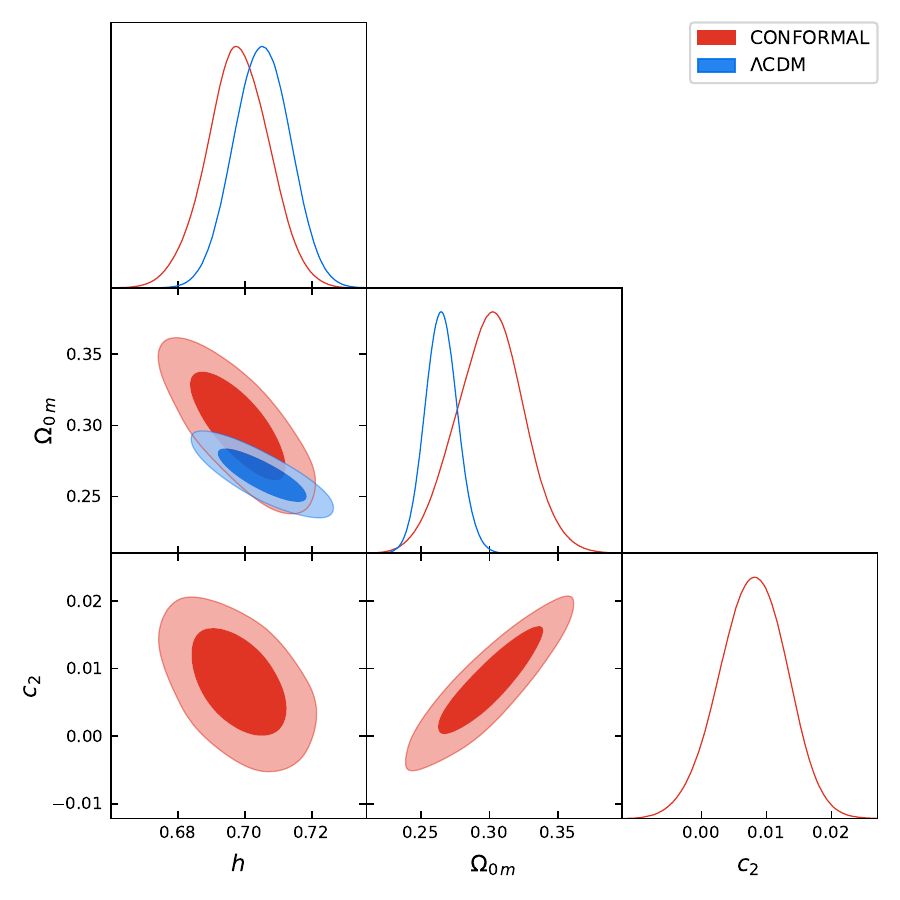}
\caption{Joint constraints of $h$, $\Omega_{0\,m}$, and $c_{2}$ for the flat $\Lambda$CDM model and the Conformal scenario, using Bayesian statistical analysis of Sec.~\ref{cosmo-constraints}. The admissible regions correspond to $1\sigma\left(68.3\%\right)$, and $2\sigma\left(95.5\%\right)$, CL, respectively. In this case, the best-fit values of the Conformal scenario at $1\sigma$ are $h=0.698^{+0.0093}_{-0.0093}$, $\Omega_{0\,m}=0.300^{+0.025}_{-0.025}$, and $c_{2}=0.0079^{+0.0051}_{-0.0051}$.}\label{joint}
\end{figure}

Table~\ref{tab:cobaya} shows the best-fit values for the parameters of each model from the OHD and Joint (OHD + SNe Ia) analysis; together with their $\chi^{2}_{min}$, and BIC indicators. Furthermore, figures~\ref{OHD} (OHD) and~\ref{joint} (Joint) show the posteriors of the parameters within the scenarios $\Lambda$CDM (blue) and Conformal (red). The best-fit values for the Conformal model, derived from the joint analysis, are $h=0.698^{+0.0093}_{-0.0093}$, $\Omega_{0\,m}=0.300^{+0.025}_{-0.025}$, and $c_{2}=0.0079^{+0.0051}_{-0.0051}$. The relations between the parameters $\Omega_{0\,m}$ and $h$ are quite similar to Hern\'{a}ndez-Jim\'{e}nez, et. al.~\cite{Hernandez-Jimenez:2022daw}; however, their framework, although it presents analogous theoretical backgrounds, the authors analysed a barotropic cosmological parameter as an ansatz. Moreover, these upshots are also reported by~\cite{Corral:2020lxt,LinaresCedeno:2020uxx}. Furthermore, the joint Conformal outcome indicates a lower $\chi^{2}_{min}$ than $\Lambda$CDM; nevertheless, given that the latter scenario has fewer free parameters, it gives a lower BIC. Indeed, we obtain $\Delta BIC_{Joint}=5.24$. Although this value implies `positive' evidence in favour of $\Lambda$CDM, there is still room to constrain the parameter space further with additional early and late times data. 

\begin{table}[h!]
\centering
\begin{tabular}{| c | c | c |}
\hline
Data  & \qquad Best-fit values: mean$^{+1\sigma}_{-1\sigma}$ \qquad  & Goodness of fit \\
\hline
 &  $h \qquad\qquad \Omega_{0\, m} \qquad\qquad c_{2}$  & $\chi^{2}_{min}$ \quad BIC \\
\hline
\multicolumn{3}{ |c| }{$\Lambda$CDM} \\ 
\hline
OHD  & $0.715^{+0.0095}_{-0.0095} \quad 0.247^{+0.014}_{-0.014} \quad -- $  & $28.89 \quad 36.79 $\\
\hline
Joint  & $0.7052^{+0.0082}_{-0.0082} \quad 0.265^{+0.012}_{-0.012} \quad -- $  & $1059.59 \quad 1073.60 $\\
\hline
\multicolumn{3}{ |c| }{Conformal} \\ 
\hline
OHD  & $0.714^{+0.014}_{-0.014} \quad 0.252^{+0.041}_{-0.046} \quad 0.0008^{+0.0071}_{-0.0080} $ & $28.86 \quad 40.71  $ \\
\hline
Joint  & $0.698^{+0.0093}_{-0.0093} \quad 0.300^{+0.025}_{-0.025} \quad 0.0079^{+0.0051}_{-0.0051} $ & $1057.83 \quad 1078.84  $ \\
\hline
\end{tabular}
\caption{Results of the best-fit parameters and statistical indicators. The uncertainties correspond to $1\sigma$ $(68.3\%)$ of the confidence level (CL). We consider the criterion $R-1<1.0\times 10^{-2}$ for the OHD data and the joint analysis.}
\label{tab:cobaya}
\end{table}

To end this section, we want to stress that the joint analysis of Conformal produces a smaller $H_{0}^{\rm Conformal}=69.80\rm\,\, Km \,s^{-1}\,Mpc^{-1}$ ($H_{0}=100\,h\rm\,\, Km \,s^{-1}\,Mpc^{-1}$) in contrast to the flat $\Lambda$CDM result $H_{0}^{\Lambda\rm CDM}=70.52\rm\,\, Km \,s^{-1}\,Mpc^{-1}$. In fact, this outcome closes the breach with respect to the CMB value $H_{0}^{\rm CMB}=67.70\rm\,\, Km \,s^{-1}\,Mpc^{-1}$ (Planck + BAO) \cite{Planck:2018nkj}.  


However, this encouraging result comes with shortcomings. As we have shown earlier in Sect.~\ref{cosmological_model}, the evident divergence at $\Xi^{2}=2$ occurs soon enough that this cannot be screened by the other energy components, leading to a universe completely dominated by matter $\rho_{m}$ in the past. Despite that, we expect that more statistical analysis, using CMB data, is needed to make a more robust study.

%
%
\section{Conclusions}\label{conclusions}

By proposing that metric perturbations are associated with conformal transformations of the metric tensor and imposing that the Einstein equations must be satisfied by the perturbed quantities, we derived a functional form for the cosmological parameter on the background manifold in terms of its counterpart on the extended manifold. This approach differs from other formulations within the framework of Relativistic Quantum Geometry, where the functional form is typically introduced ad hoc~\cite{Bellini:2024meo,Hernandez-Jimenez:2022daw} or motivated by imposing a specific normalisation condition on the velocities of relativistic observers~\cite{Bellini:2025snh}. Although beyond the scope of the present work, it would be of interest to investigate which types of perturbative schemes, following the strategy of imposing the Einstein equations on the extended manifold, may lead to functional forms of the cosmological parameter consistent with those proposed in other works in the literature.

By statistical analysis, we calculated the best-fit value of $\Theta =\{h, \Omega_{0\, m}, c_{2} \}$ using the affine-invariant MCMC. We used late-time data for OHD and SNe Ia. The result of the joint analysis is presented in Table~\ref{tab:cobaya}. For the joint data, we have: $h=0.698^{+0.0093}_{-0.0093}$, $\Omega_{0\,m}=0.300^{+0.025}_{-0.025}$, and $c_{2}=0.0079^{+0.0051}_{-0.0051}$. Moreover, the joint Conformal outcome indicates a lower $\chi^{2}_{min}$ than $\Lambda$CDM; nevertheless, given that the latter scenario has fewer free parameters, it gives a lower BIC. In fact, we obtain $\Delta BIC_{Joint}=5.24$. Although this value implies `positive' evidence in favour of $\Lambda$CDM, there is still room to constrain the parameter space further with additional early and late times data. Despite the attractive outcome, limitations exist. Section~\ref{cosmological_model} highlights the significant divergence at $\Xi^{2}=2$, which occurs too early to be mitigated by other energy components, resulting in a past universe ruled by matter $\rho_{m}$, see figs.~\ref{fig:Omegas} and~\ref{fig:q}. Thus, the larger $z$ the divergence occurs, this minimally impacts the relevant parameters. However, further statistical analysis incorporating CMB data is necessary for a more definitive investigation.  

Considering the aforementioned shortcomings, in the appendix~\ref{appendix_a} we present the same Bayesian statistical analysis as in Sec.~\ref{cosmo-constraints}; however, we fixed the value of the constant $c_{2}=0.0029$, since this value represents the highest numerical estimate, assuming that $\Omega(z)$ and $q(z)$ comply with our past expectations of the universe. The best-fit values for the Conformal$^{\rm OHD}$ model are $h=0.7103^{+0.0092}_{-0.0092}$, $\Omega_{0\,m}=0.264^{+0.013}_{-0.015}$; and the Conformal$^{\rm Joint}$ model are $h=0.7006^{+0.0093}_{-0.0093}$, $\Omega_{0\,m}=0.281^{+0.013}_{-0.013}$. Considering that each scheme has two free parameters, it is pertinent to use the $\chi^{2}_{min}$ values for comparison. Therefore, the Conformal$^{\rm Joint}$ result denotes better compatibility between the model and the data, as its $\chi^{2}_{min}=1059.07$ is slightly smaller than that of $\Lambda$CDM $\chi^{2}_{min}=1059.59$. However, this moderate improvement does not definitively indicate that the Conformal model is above the $\Lambda$CDM framework. Moreover, in contrast to Sec.~\ref{cosmological_model} analysis, this time the two Conformal cases exhibit completely different evolutions for $z\gg 1$. Only the Conformal$^{\rm Joint}$ outcome exhibit a regular path; whilst the Conformal$^{\rm OHD}$ instance does the opposite. This behaviour might be explained by a lower density parameter value $\Omega_{0\,m}^{(\rm OHD)}=0.264$, which, in turn, produces a larger $h^{(\rm OHD)}=0.7103$. So, we suspect that these two parameters modify the normalised Hubble $E^{2}(z)$, producing different results. Ultimately, the location of the divergence is not the only sign of the abrupt departure from our must-likely history of the universe; the other free parameters are also relevant. 

In conclusion, we highlight some limitations of the present study. First, the restriction of metric perturbations to conformal deviations, as given by eq.~\eqref{perturbacionesg}, limits the generality of the possible functional forms that the cosmological parameter may acquire. A further limitation lies in the arbitrariness of the choice of $\bar{\Lambda}$ used to obtain eq.~\eqref{onda3}, which was primarily made to arrive at a simpler expression for the cosmological parameter. For instance, one could have assumed $\bar{\Lambda} = \Lambda_0$, in which case $\Lambda$ would include an additional term $\Lambda_T$; alternatively, $\bar{\Lambda}$ could have been determined from a conservation equation satisfied by $\bar{T}_{\alpha\beta}$. 

Strengthening the theoretical foundations underlying the derivation of functional forms for the cosmological parameter, along with fitting the corresponding free parameters using larger observational data sets, may produce models that are statistically favoured over \(\Lambda\)CDM. This topic has attracted significant attention in recent years and remains the subject of ongoing research.

%
%

\appendix
\section{Fixed constant $c_{2}=0.0029$ scenario}\label{appendix_a}
In this appendix, we include the same Bayesian statistical analysis as in Sec.~\ref{cosmo-constraints}; however, we fixed the value of the constant $c_{2}=0.0029$. This value represents the highest numerical estimate, assuming that $\Omega(z)$ and $q(z)$ comply with our historical expectations of the universe. We present table~\ref{tab:appendix} with our results.  
\begin{table}[h!]
\centering
\begin{tabular}{| c | c | c |}
\hline
Data  & \qquad Best-fit values: mean$^{+1\sigma}_{-1\sigma}$ \qquad  & Goodness of fit \\
\hline
 &  $h \qquad\qquad \Omega_{0\, m}$ & $\chi^{2}_{min}$ \\
\hline
\multicolumn{3}{ |c| }{$\Lambda$CDM} \\ 
\hline
OHD  & $0.715^{+0.0095}_{-0.0095} \quad 0.247^{+0.014}_{-0.014}$  & $28.89 $\\
\hline
Joint  & $0.7052^{+0.0082}_{-0.0082} \quad 0.265^{+0.012}_{-0.012}$  & $1059.59 $\\
\hline
\multicolumn{3}{ |c| }{Conformal} \\ 
\hline
OHD  & $0.7103^{+0.0092}_{-0.0092} \quad 0.264^{+0.013}_{-0.015} $ & $28.90 $ \\
\hline
Joint  & $0.7006^{+0.0093}_{-0.0093} \quad 0.281^{+0.013}_{-0.013} $ & $1059.07 $ \\
\hline
\end{tabular}
\caption{Results of the best-fit parameters and statistical indicators. The uncertainties correspond to $1\sigma$ $(68.3\%)$ of the confidence level (CL). We consider the criterion $R-1<1.0\times 10^{-2}$ for the OHD data and the Joint analysis. Here we fix $c_{2}=0.0029$.}
\label{tab:appendix}
\end{table}

\begin{figure}[htbp] 
\includegraphics[scale=0.95]{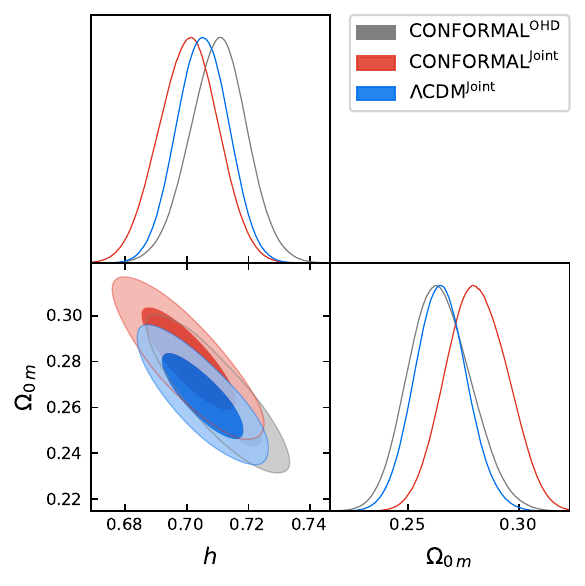}
\caption{OHD and Joint data sets constraints of $h$ and $\Omega_{0\,m}$, and fixed $c_{2}=0.0029$, for the flat $\Lambda$CDM model and the Conformal scenario, using Bayesian statistical analysis of Sect.~\ref{cosmo-constraints}. The admissible regions correspond to $1\sigma\left(68.3\%\right)$, and $2\sigma\left(95.5\%\right)$, CL, respectively. In this case, the best-fit values of the Conformal$^{\rm Joint}$ scenario at $1\sigma$ are $h=0.7006^{+0.0093}_{-0.0093}$ and  $\Omega_{0\,m}=0.281^{+0.013}_{-0.013}$.}\label{appendix_triangle}
\end{figure}

Moreover, fig.~\ref{appendix_triangle} shows the posteriors of the parameters within the scenarios $\Lambda$CDM (blue), Conformal$^{\rm OHD}$ (grey), and Conformal$^{\rm Joint}$ (red). The best-fit values for the Conformal$^{\rm Joint}$ model are $h=0.7006^{+0.0093}_{-0.0093}$, $\Omega_{0\,m}=0.281^{+0.013}_{-0.013}$. This time, since all schemes have two free parameters, it becomes relevant to compare them only using the values $\chi^{2}_{min}$. Consequently, the Conformal$^{\rm Joint}$ outcome implies improved model-data compatibility. Nevertheless, a marginally improved outcome is not a robust indication that the Conformal framework is favoured over the $\Lambda$CDM paradigm.

\begin{figure}[htbp] 
\includegraphics[scale=0.75]{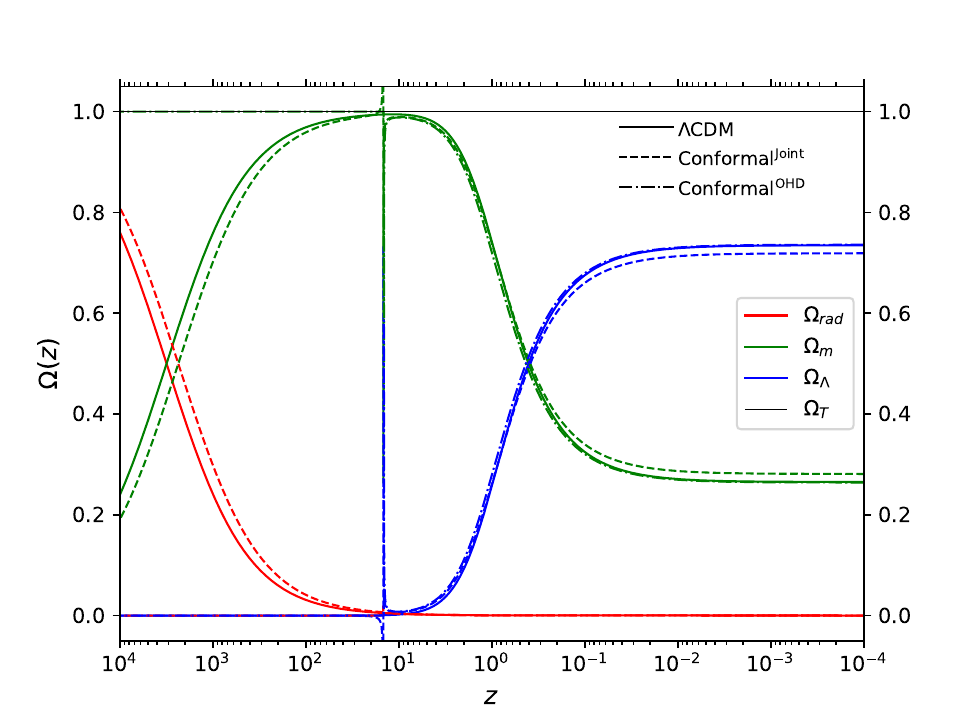}
\caption{Cosmological evolution of $\Omega_{\Lambda}(z)$ (blue), $\Omega_{m}(z)$ (green), $\Omega_{rad}(z)$ (red), and $\Omega_{T}(z)=\sum_{i}\Omega_{i} = 1$ (black). Solid lines correspond to ($\Lambda$CDM) case; dashed ones for Conformal$^{\rm Joint}$ (OHD + SNe Ia) model; and dashed dots Conformal$^{\rm OHD}$ (Observational Hubble Data). Both vertical cuts represent the divergence for $\Xi^{2}=2$ at $z\sim 16$.}\label{fig:Omegas_appendix}
\end{figure}
\begin{figure}[htbp] 
\includegraphics[scale=0.75]{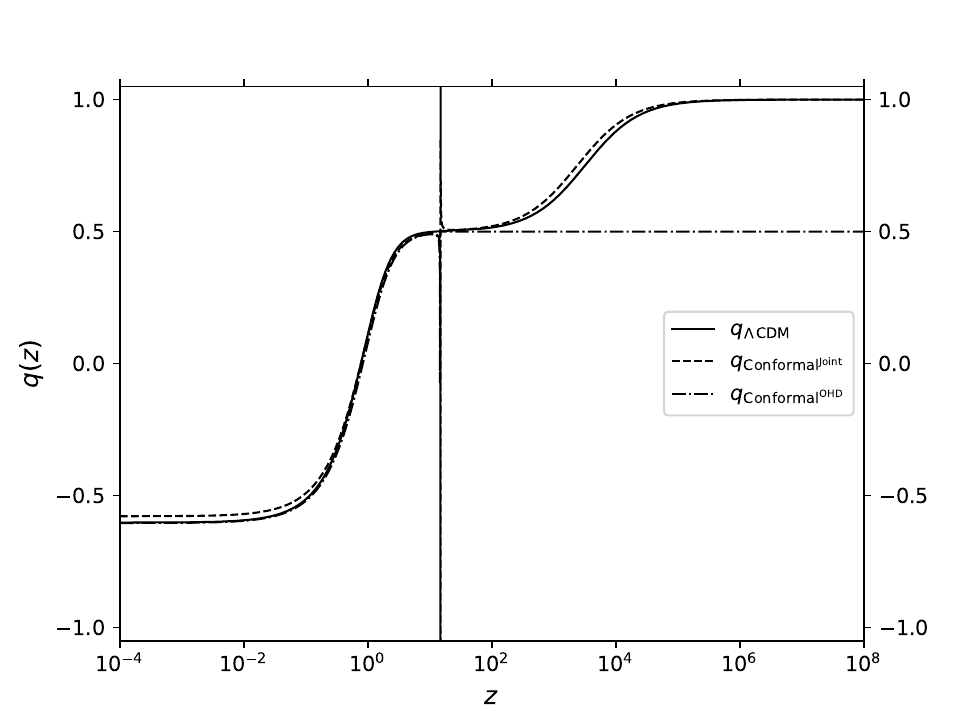}
\caption{Evolution of the deceleration parameter $q(z)$ in terms of the redshift $z$. Solid lines correspond to $\Lambda$CDM framework; then dashed ones for Conformal$^{\rm Joint}$ (OHD + SNe Ia) model; and dashed dots Conformal$^{\rm OHD}$ (Observational Hubble Data). Both vertical cuts represent the divergence for $\Xi^{2}=2$ at $z\sim 16$.}\label{fig:q_appendix}
\end{figure}

Finally, note that in both figs.~\ref{fig:Omegas_appendix} and~\ref{fig:q_appendix} there are two abrupt vertical lines, where the divergences occur, i.e. at $\Xi^{2} = 2$. In contrast to Sec.~\ref{cosmological_model} analysis, this time the two Conformal cases exhibit completely different evolutions for $z\gg 1$. In fact, although both singularities occur around $z\sim 16$, only the Conformal$^{\rm Joint}$ outcome exhibit a regular path; whilst the Conformal$^{\rm OHD}$ instance does the opposite, its universe is ruled only the matter component ($\rho_{m}$), right after its divergence, leaving all other elements absolutely erased. This behaviour might be explained by a lower density parameter value $\Omega_{0\,m}^{(\rm OHD)}=0.264$ compared to $\Omega_{0\,m}^{(\rm Joint)}=0.281$; and this, in turn, yields a larger $h^{(\rm OHD)}=0.7103$. These two parameters modify the normalised Hubble $E^{2}(z)$, so this does not evolve accordingly to our expected past. In conclusion, the point of the divergence is not the only indicative of the abrupt departure from our probably past.

\acknowledgments 

This work was supported by the CONAHCYT Network Project No. 376127 {\it Sombras, lentes y ondas gravitatorias generadas por objetos compactos astrofísicos}. J.I.M. is supported by CONICET, Argentina (PIP 11220200100110CO) and UNMdP (EXA1156/24). R.H.J is supported by CONAHCYT Estancias Posdoctorales por M\'{e}xico, Modalidad 1: Estancia Posdoctoral Acad\'{e}mica and by SNI-CONAHCYT. C.M. wants to thank SNI-CONAHCYT, PROINPEP-UDG, PROSNI-UDG and SEP-PRODEP.

%
%
\bibliographystyle{unsrt}
\bibliography{conformal_Lambda}

\end{document}